\documentclass[reprint,twocolumn,aps,prB,amsmath,amssymb,floatfix,superscriptaddress,longbibliography]{revtex4-1}
\usepackage{graphicx}
\usepackage{epsfig}
\usepackage{bm}
\usepackage{dcolumn}
\usepackage{color}
\usepackage{physics}
\usepackage{float}
\usepackage[colorlinks,urlcolor=blue,citecolor=blue,linkcolor=magenta]{hyperref}
\makeatletter
\newcommand*{\rom}[1]{\expandafter\@slowromancap\romannumeral #1@}
\makeatother
\usepackage{tikz}
\usepackage{subfigure}

\begin{document}

\preprint{APS/123-QED}
\preprint{This line only printed with preprint option}

\title{Emergent multi-loop nested point gap in a non-Hermitian quasiperiodic lattice}

\author{Yi-Qi Zheng}
\affiliation{School of Physics, South China Normal University, Guangzhou 510006, China}

\author{Shan-Zhong Li}
\email[Corresponding author: ]{szhongli@m.scnu.edu.cn}
\affiliation{School of Physics, South China Normal University, Guangzhou 510006, China}

\author{Zhi Li}
\email[Corresponding author: ]{lizphys@m.scnu.edu.cn}
\affiliation {Key Laboratory of Atomic and Subatomic Structure and Quantum Control (Ministry of Education), Guangdong Basic Research Center of Excellence for Structure and Fundamental Interactions of Matter, School of Physics, South China Normal University, Guangzhou 510006, China}

\affiliation {Guangdong Provincial Key Laboratory of Quantum Engineering and Quantum Materials, Guangdong-Hong Kong Joint Laboratory of Quantum Matter, Frontier Research Institute for Physics, South China Normal University, Guangzhou 510006, China}

\date{\today}

\begin{abstract}
We propose a geometric series modulated non-Hermitian quasiperiodic lattice model, and explore its localization and topological properties. The results show that with the ever-increasing summation terms of the geometric series, multiple mobility edges and non-Hermitian point gaps with high winding number can be induced in the system. The point gap spectrum of the system has a multi-loop nested structure in the complex plane, resulting in a high winding number. In addition, we analyze the limit case of summation of infinite terms. The results show that the mobility edges merge together as only one mobility edge when summation terms are pushed to the limit. Meanwhile, the corresponding point gaps are merged into a ring with winding number equal to one. Through Avila's global theory, we give an analytical expression for mobility edges in the limit of infinite summation, reconfirming that mobility edges and point gaps do merge and will result in a winding number that is indeed equal to one.
\end{abstract}

\maketitle

\section{Introduction}
Anderson localization reveals the phenomenon of how electron Bloch wave diffusion is suppressed by random disorder~\cite{PWAnderson1958}. To this day, the discussion of localization problems remains a dynamic field in condensed matter physics. The scaling theory established in the middle of last century fully reveals that Anderson localization caused by disorder is dependent on system dimension~\cite{EAbrahams1979,PALee1985,FEvers2008,BHetenyi2021}. Specifically, in low dimensions ($D<3$), arbitrarily small disorder suffices to localize the system~\cite{EAbrahams1979}. For the case of $D=3$, extended and localized states can coexist, separated by the so called mobility edge (ME)~\cite{FEvers2008,ALagendijk2009,NFMott1967,JKoo1975}. This implies that, the localization phase transition depends on the disorder strength and energy in 3D system.

Different from random disorder, quasiperiodic systems, which are intermediate between disorder and order system, are widely used to study localization phase transitions and MEs in low-dimensional systems. A widely known 1D quasiperiodic system is called the Aubry-Andr{\'e} (AA) model, whose transition point from extended to localized phases can be exactly determined by self-duality conditions~\cite{SAubry1980,SYJitomirskaya1999}. The self-duality condition also predicts that the extended and localized states cannot coexist and that there are no MEs. Consequently, generalized versions of the AA model have been proposed to study MEs, including modulated lattice structures, modified quasiperiodic potentials~\cite{APadhan2022,SDasSarma1988,SDasSarma1990,TLiu2022,SGaneshan2015,XLi2017,HYao2019,XLi2020,YCZhang2022,XCZhou2023,YWang2020,XPLi2016,ZLu2022,SRoy2021}, and the introduction of long-range hopping~\cite{JBiddle2011,XXia2022,XDeng2019,MGoncalves2023,JBiddle2010}. Beyond these, the AA model has found widespread applications in exploring topological Anderson insulators~\cite{JLi2009,GQZhang2021,DWZhang2020b,DWZhang2020c}, transport properties~\cite{YEKraus2012,JCCharlier2007,SHuang2024}, entanglement transitions~\cite{LZhou2024,LZhou2024a,LZhou2023,SZLi2024}, and multifractal phases~\cite{XBeng2019,MSarkar2022,SZLi2023,XLin2023,HYao2019,MGoncalves2023,FLiu2015,XCZhou2023}. These remarkable properties can be experimentally realized using cold atomic systems~\cite{FAAn2018,DTanese2014,GModugno2010,MLohse2016,SNakajima2016,DWZhang2018}, optical lattices~\cite{FAAn2018,HPLuschen2018,SAGredeskul1989,DNChristodoulides2003,YEKraus2012,TPertsch2004,HPL2018,MSchreiber2015,MLohse2016,SNakajima2016}, photonic crystals~\cite{YLahini2009,MVerbin2013,MVerbi2015}, and superconducting qubits~\cite{PRoushan2017,HLi2023}.

Recently, there has been a growing interest in localization and topological phase transitions in non-Hermitian quasiperiodic systems~\cite{PHe2022,DSBorgnia2023,DSBorgnia2023a,SZLi2024b,SZLi2024a,APadhan2024,DWZhang2020b,HJiang2019,SLonghi2019,YLiu2020,QBZeng2020a,TLiu2020,XCai2021,YLiu2021a,JClaes2021,LZTang2021,SLonghi2021a,SLonghi2021,LZhou2022,XCai2022,WHan2022,QLin2022,YLiu2021,LJZhai2020,LZTang2022,TQian2024,YLiu2021c,QBZeng2020b}, where non-Hermiticity can stem from non-reciprocal hopping or complex on-site potential. Essentially, the two types of non-Hermitian modulations are related. One can uncover this relation through the dual transformation, that complex on-site potential is equivalent to non-reciprocal hopping in the dual space, vice versa. Furthermore, when there are MEs in the system, the above two non-Hermitian modulation will exhibit the dual relationship. Specifically, under the condition of complex on-site potential (non-reciprocal hopping) modulation, the extended states (localized states) possess  the real spectra, while the localized states (extended states) correspond to the complex spectra. Such complex spectra form closed loops in the complex plane and cause the system to exhibit non-zero topological winding numbers~\cite{ZGong2018,KZhang2020,PHe2020}. The Anderson transition point in energy coincides with the PT symmetry breaking~\cite{LWZhou2022,SLonghi2019,TLiu2022a,XXia2022,SLonghi2021,YLiu2021c,YLiu2021,YLiu2020,YLiu2021a}.

So far, although a lot of efforts have been made in the study of the traditional point gap with winding number equal to one~\cite{SLonghi2019,QBZeng2020b,KKawabata2019,SLonghi2019a}, few papers discuss the point gap with higher-spectrum winding numbers~\cite{YLiu2021,APadhan2024}. This work, therefore, is devoted to studying the non-Hermitian topological point gap with higher-spectrum winding numbers, the corresponding localization properties and MEs.

The main findings of this manuscript are as follows. \textit{Multiple MEs and multi-loop nested higher-spectrum point gap can emerge in  non-Hermitian quasiperiodic lattice system.}

The rest of the paper is structured as follows. In Sec.~\ref{Sec.2}, we give a brief introduction to the theoretical model and the main findings. In Sec.~\ref{Sec.3}, through three cases, we provide the evidence of the emergent multi-loop nested point gaps. By Avila's global theory, we analytically discuss the case of infinity summation limit in Sec.~\ref{Sec.4}.  In Sec.~\ref{dualspace}, we discuss the physical meaning of non-Hermitian point gap topologies. We summarize this manuscript in Sec.~\ref{Sec.6}.


\section{Model}\label{Sec.2}
\begin{figure}[b]
\centering
\includegraphics[width=8.5cm]{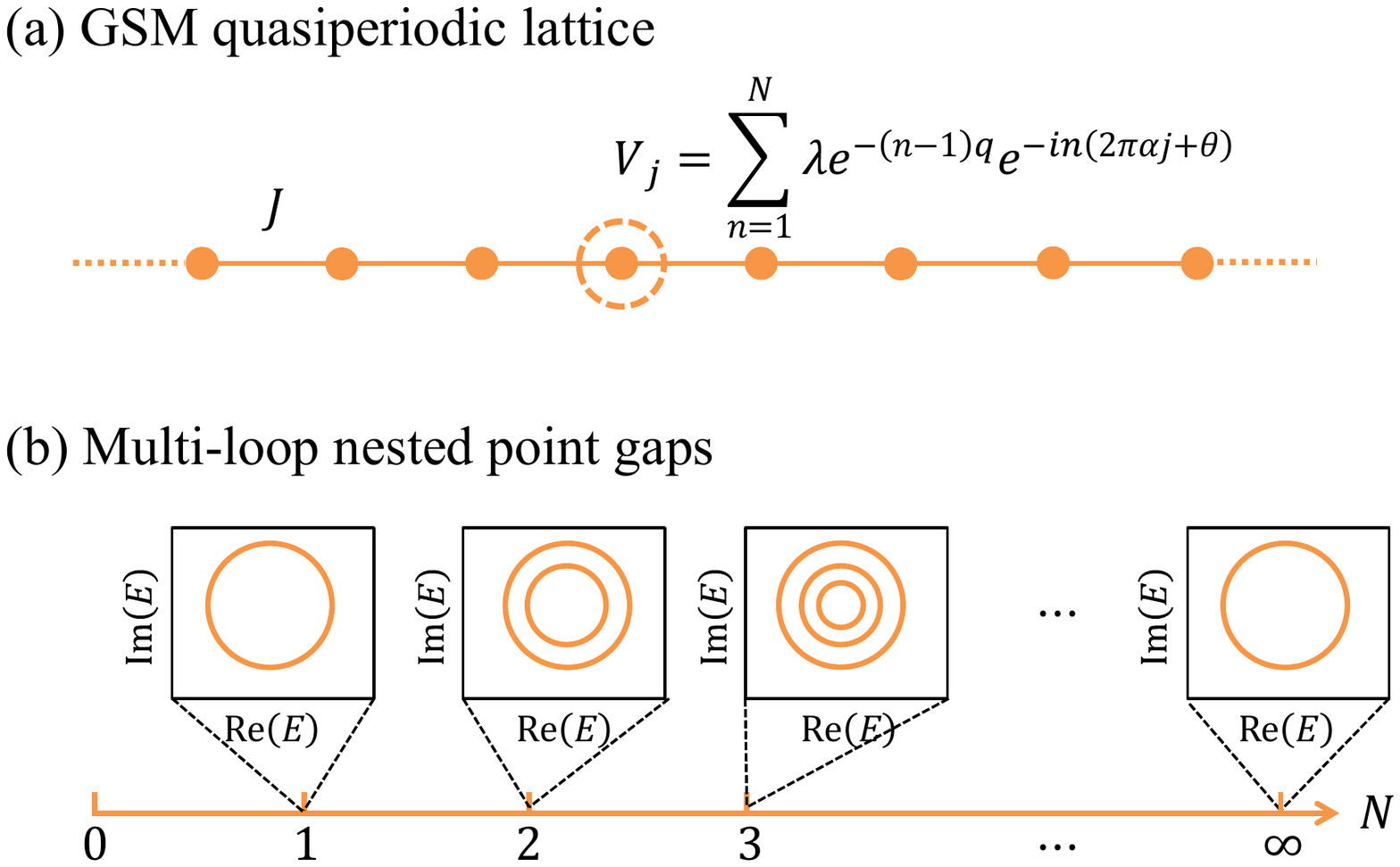}
\caption{The scheme diagram of GSM lattice model (a) and multi-loop nested point gaps spectra (b).}
    \label{F0}
\end{figure}

Let's start at a non-Hermitian geometric series modulated (GSM) quasiperiodic potential~[see Fig.~\ref{F0}(a)]. The corresponding Hamiltonian reads
\begin{equation}\label{Hami}
H=\sum_{j=1}^{L-1}(Jc_{j}^{\dagger}c_{j+1}+\mathrm{H.c.})+\sum_{j=1}^{L}V_{j}c_{j}^{\dagger}c_{j},
\end{equation}
where
\begin{equation}\label{Vj}
V_{j}=\sum_{n=1}^{N}{\lambda}e^{-(n-1)q}e^{in(2{\pi}{\alpha}j+{\theta})},
\end{equation}
and $c_{j}^{\dagger}$ ($c_{j}$) denotes the fermion creation (annihilation) operator on site $j$. $J$ represents the nearest-neighboring hopping strength and $L$ is the system size. $V_{j}$ is the GSM potential with quasiperiodic strength $\lambda$. $N$ is the number of summation terms of the geometric series. The corresponding common ratio of the geometric series is $e^{-q}e^{i(2\pi\alpha j+ \theta)}$, where $\alpha$ is an irrational number to guarantee that the potential energy has a quasiperiodic structure and $\theta$ is the phase offset. 
When $\theta=0$, one has $V_{-j}=V_{j}^{\ast}$ and the system is PT symmetric in rigorous periodic boundary conditions (PBCs). Without loss of generality, in numerical computation, we set $J=1$ as unit energy, $\theta=0$, $q=0.2$ and $\alpha=\lim_{m\rightarrow\infty}F_{m-1}/F_{m}=(\sqrt{5}-1)/2$ with $F_{m}$ being the $m$-th Fibonacci number. For finite size cases, we set system size $L=F_{m}$ and $\alpha=F_{m-1}/F_{m}$ to ensure rigorous PBCs.

We focus on the properties of MEs and topology point gaps. To this end, we calculate two key quantities. 

One is the fractal dimension, which constitutes the core quantity describing the localization properties and MEs. The fractal dimension can be defined as~\cite{YWang2020}
\begin{equation}\label{Gamma}
\Gamma_m=-\lim_{L\rightarrow\infty}\frac{\ln\xi_m}{\ln L},
\end{equation}
where $\xi_m=\sum_{j=1}^{L}|\psi_{j,m}|^4$ is the corresponding inverse participation ratio. $\psi_{j,m}$ denotes the amplitude for the $m$-th eigenstate at the $j$-th site. For the localized (extended) state, $\Gamma=0$ ($\Gamma=1$), while the critical state corresponds to $\Gamma\in(0,1)$. Note that, $\Gamma$ tending to $1(0)$ would indicate the extended (localized) state at finite size. 

The other is the winding number, which represents a key quantity that reflects the topological properties and can be defined as
\begin{equation}\label{omega}
\omega(E_{b})=\lim_{L\rightarrow\infty}\frac{1}{2\pi i}\int_{0}^{2\pi} d\theta\frac{\partial}{\partial \theta} \ln\left\{\mathrm{ det}[H(\theta\rightarrow\frac{\theta}{L})-E_{b}]\right\},
\end{equation}
which describes the number of times the spectrum revolving around the base point $E_{b}$ in the complex plane~\cite{ZGong2018,SLonghi2019}. Below, we exhibit that multi-loop nested point gaps~[see Fig.~\ref{F0}(b)] emerge in the lattice model~\eqref{Hami}.

\section{Multi-loop nested point gaps}\label{Sec.3}

\begin{figure*}[thbp]
\centering
\includegraphics[width=15cm]{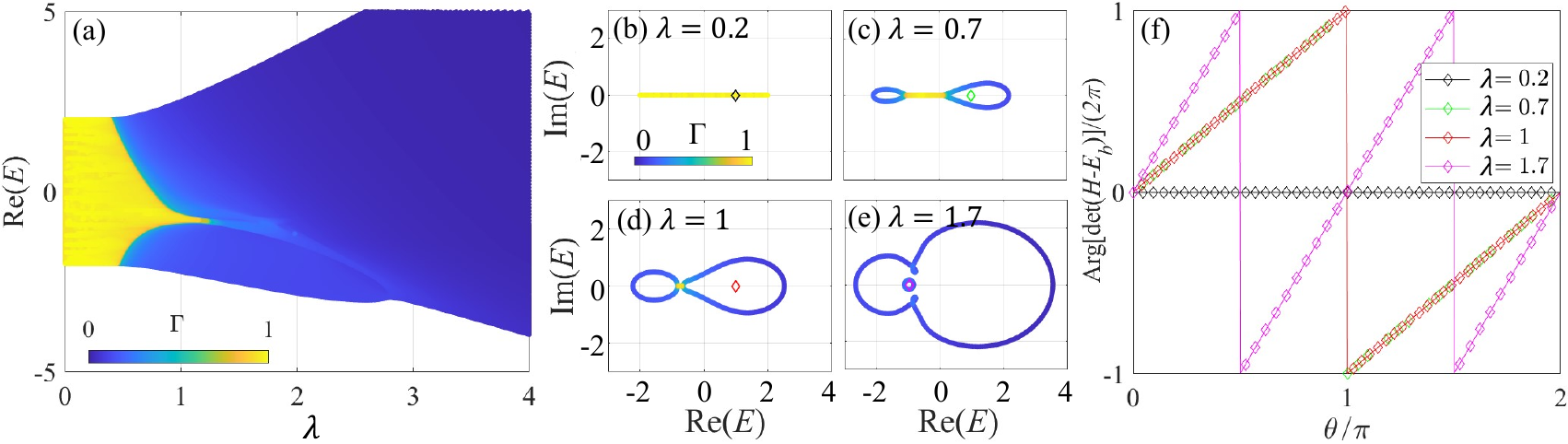}
\caption{(a) The phase diagram of fractal dimension ${\Gamma}$ in the $\lambda-E$ plane. Eigenenergies in the complex plane with $\lambda=0.2$ (b), $\lambda=0.7$ (c), $\lambda=1$ (d) and $\lambda=1.7$ (e). (f) Arg[det($H-E_{b}$)] versus $\theta$, where we set $E_{b}=1,~1,~1,~-0.95$ [marked as rhombus in (b)-(e)]. Throughout, we set $L=610$, $q=0.2$,~$\alpha=377/610$ and $N =2$.}
    \label{F1}
\end{figure*}

\subsection{Traditional point gap with $\omega=1$}
First, we consider the case of $N = 1$. The quasiperiodic potential contains only the first term of the geometric series~\eqref{Vj}, that is,
\begin{equation}
 V_{j}=\lambda e^{i(2\pi\alpha j+\theta)}.
\end{equation}
Under such circumstance, the model is an integrable system. The critical point of Anderson transition is located at $\lambda_c=J$. The corresponding Lyapunov exponent $\gamma=\ln(\lambda/J)$~\cite{SLonghi2019,SLonghi2019a,YLiu2021}. When $\lambda<J$ ($\lambda>J$), the system is extended (localized) and the eigenenergy spectrum $E(k)=2\cos(k)$ [$E(k)=\lambda e^{ik}+\lambda e^{-ik}$] with $k\in(0,2\pi]$, which corresponds to the trivial (topological) phase. In the topological phase, the maximum winding number $\omega=1$, which equals to the summation number of the GSM potential.

\subsection{The point gap with $\omega=2$}
Now, let's turn to point gaps with high winding numbers. Under the condition of $N=2$, the corresponding quasiperiodic potential reads
\begin{equation}\label{Vj2}V_{j}={\lambda}e^{i(2{\pi}{\alpha}j+{\theta})}+\lambda e^{-q}e^{2i(2{\pi}{\alpha} {j}+\theta)}.
\end{equation}
In Fig.~\ref{F1}(a) we exhibit the fractal dimensions $\Gamma$ versus $\lambda$ and $E$. By contrast to the case of traditional point gap, the multi-loop nested point gaps emerge here accompanied by MEs. In concrete terms, for the former, there is no ME, and only a clean critical point ($\lambda = J$) exists between the localized and extended states. For the latter, however, the potential of one more summation term cause the ME to be in the interval $0.48 < \lambda < 1.2$~[see Fig.~\ref{F1}(a)]. Furthermore, the localized states gradually invade the extended region until all eigenstates become localized with an increasing $\lambda$. Meanwhile, as the eigenspectrum changes from pure real to complex, the system's PT symmetry is gradually broken~[see Fig.~\ref{F1}(b)-(e)]. One can also see that the localized state forms two separated circles at the maximum and minimum eigenvalues, which are connected by the eigenvalues of the extended state. And when all the eigenstates are localized, the two circles fuse to form a multi-loop nested structure~[see Fig.~\ref{F1}(e)]. Fig.~\ref{F1}(f) exhibits the argument of the determinants $H-E_{b}$ as a function of the flux $\theta$, where the corresponding base points $E_{b}$ are marked as a rhombus in Fig.~\ref{F1}(b)-(e). One can find that the Arg[det($H-E_b$)] is always zero with $\theta$ for the extended phase ($\lambda = 0.2$), corresponding to the trivial phase with winding number $\omega = 0$. Under the condition of $\lambda=0.7$ and $\lambda=1$, the Arg[det($H-E_b$)] forms a cycle with the $\theta$, corresponding to the topological phase with winding numbers $\omega=1$. Under the condition of complete localization ($\lambda=1.7$), a topological phase with high winding number $\omega=2$ emerges with an increasing $\theta$. 

In other words, two stepwise phase transitions occur in the system, i.e., localization and topological phase transition. These two phase transitions have exactly identical critical points and corresponding regions of different phases. That is to say, when the system gradually changes from the extended phase, through the intermediate phase, and finally into the localized phase, the topological properties of the system also gradually change from the trivial phase, through the topological phase of $\omega=1$, and finally into the topological phase of $\omega=2$. Meanwhile, the symmetry of the system will start with PT symmetry, then enter a partial PT symmetry broken region, and finally become a state where PT symmetry is completely broken.

\begin{figure*}[tbp]
\centering
\includegraphics[width=15cm]{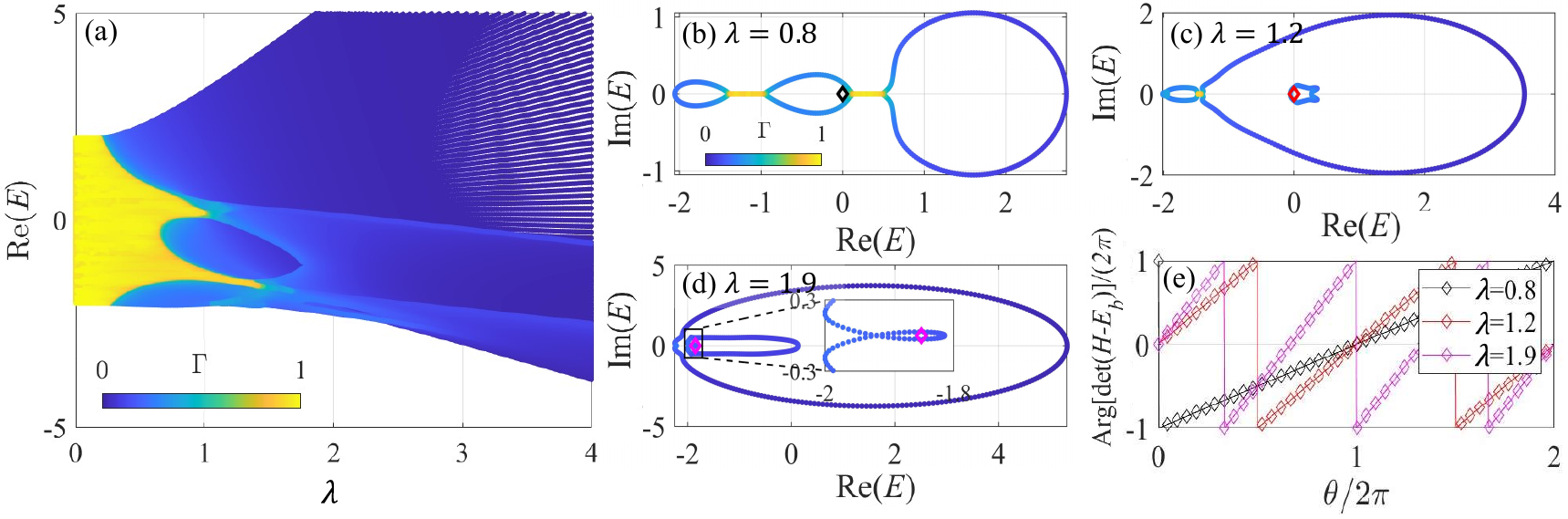}
\caption{(a) The phase diagram of fractal dimension ${\Gamma}$ in the $\lambda-E$ plane. Eigenenergies in the complex plane with $\lambda=0.8$ (b), $\lambda=1.2$ (c), $\lambda=1.9$ (d). (f) Arg[det($H-E_{b}$)] versus $\theta$, where we set $E_{b}=0,~0,~-1.85$ [marked as rhombus in (b)-(d)]. The other parameters $L=610$, $\alpha=377/610$, $q=0.2$ and $N =3$.}
    \label{F2}
\end{figure*}

\subsection{The point gap with $\omega=3$}

For the case of $N = 3$, one more term is added into the summation of GSM potential, which can be written as
\begin{equation}\label{Vj3}
V_{j}={\lambda}e^{i(2{\pi}{\alpha}j+{\theta})}+{\lambda}e^{-q}e^{2i(2{\pi}{\alpha}j+{\theta})}+{\lambda}e^{-2q}e^{3i(2{\pi}{\alpha}j+{\theta})}.
\end{equation}
The corresponding fractal dimension $\Gamma$ of all eigenstates as a function of $\lambda$ is plotted in the following Fig.~\ref{F2}(a). One can find that there are more MEs and more circles of point gap spectrum~[see Fig.~\ref{F2}(b)-(d)]. Furthermore, the point gap's circles emerge one by one in the complex plane. To be more specific, two circles will appear first at either end of the real energy, and then the third circle structure appears from the middle of the energy spectrum~[see Fig.~\ref{F2}(a)]. These circle structures with complex values correspond to localized eigenstates, which are connected together in the complex plane by the pure real eigenspectrum of extended states. Later, with the increase of disorder strength $\lambda$, the three discrete circles will become increasingly larger. Finally, they merge together to form a three-layer multi-loop nested point gap~[see Fig.~\ref{F2}(b)-(d)]. The corresponding winding numbers will gradually complete the transformation of $0\rightarrow1\rightarrow2\rightarrow3$. Again, accompanied by topological and localization phase transitions, the PT symmetry gradually breaks. The corresponding arguments confirms the conclusion as shown in Fig.~\ref{F2}(e).

Combining the above results of $N=2$ and $N=3$, one can find that the winding number, the circle number of point gaps and the number of MEs are all dominated by the potential energy structure, i.e. the number of summation terms $N$. In concrete terms, the number of summation terms $N$ corresponds to the winding number $\omega=N$, the circle number of point gaps is equal to $N$, and the number of MEs is equal to $2(N-1)$.

\section{Theoretical analysis in infinite limit ($N={\infty}$)}\label{Sec.4}
Now, let's consider the limit case of $N =\infty$. Since the common ratio satisfies $|e^{-q}e^{i(2{\pi}{\alpha}j+{\theta})}|\in (0,~1]$, the corresponding geometric series converges. Then, by using the geometric series' summation formula, one can obtain
\begin{equation}\label{Vj4}V_{j}=\frac{{\lambda}e^{i(2{\pi}{\alpha}j+{\theta})}}{1-e^{-q}e^{i(2{\pi}{\alpha}j+{\theta})}}.
\end{equation}
Note that, the expression Eq.~\eqref{Vj4} is a non-Hermitian version of the Ganeshan-Pixley-Das Sarma potential~\cite{SGaneshan2015}, and the corresponding MEs can likewise be solved exactly by the Avila global theory.

\begin{figure}[tbp]
\centering
\includegraphics[width=8.5cm]{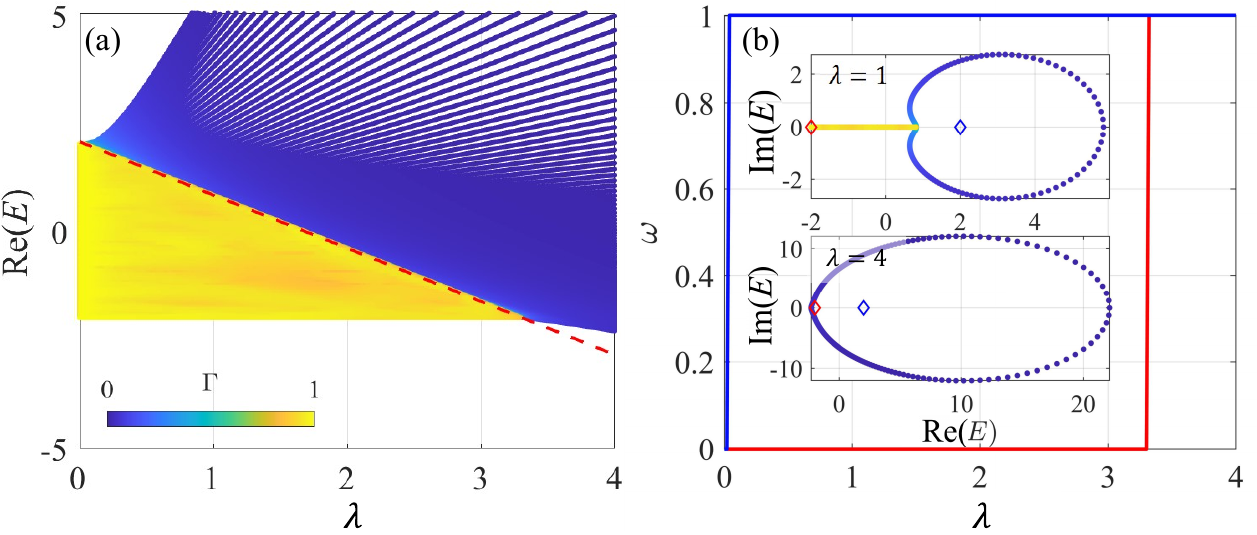}
\caption{(a) The phase diagram of fractal dimension ${\Gamma}$ in the $\lambda-E$ plane, red dashed line denotes the analytical ME. (b) The winding number versus $\lambda$ for $E_{b}=-2$ (red line) and $E_{b}=2$ (blue line). The insets exhibit point gaps in the complex plane, where the base point $E_{b}=-2,~2$ [the rhombus in the insets]. Parameters $L=610$, $\alpha=377/610$, $q=0.2$ and $N = \infty$.}
    \label{F3}
\end{figure}

Under the condition of $N=\infty$, the eigenequation for Hamiltonian~\eqref{Hami} is 
\begin{equation}
\begin{aligned}
E\psi_{j}=\psi_{j+1}+\psi_{j-1}+V_{j}\psi_{j},
\end{aligned}
\end{equation}
and the corresponding transfer matrix is
\begin{equation}\label{Phij}
T_{j}=\begin{bmatrix}
E-V_{j}  & -1\\
 1 &0
\end{bmatrix}=A_{j}B_{j},
\end{equation}
where
\begin{equation}
\begin{aligned}
&A_{j}=\frac{1}{1-e^{-q}e^{i(2{\pi}{\alpha}j+{\theta})}},\\
&B_{j}=\begin{bmatrix}
E-(Ee^{-q}+\lambda )e^{i(2\pi\alpha j+\theta)}   & -1+e^{-q}e^{i(2\pi\alpha j+\theta)} \\
1-e^{-q}e^{i(2\pi\alpha j+\theta)}  & 0
\end{bmatrix}.
\end{aligned}
\end{equation}
With the help of the transfer matrix, one can calculate the Lyapunov exponent (LE), i.e.,
\begin{equation}
\gamma=\lim_{L\rightarrow\infty}\frac{1}{L}\ln\left \| {\textstyle \prod_{j=1}^{L}}T_{j} \right \|,
\end{equation}
where $\left \|\cdot \right\|$ denotes the matrix norm. The LE can be divided into two parts, i.e., $\gamma(E)=\gamma_{A}(E)+\gamma_{B}(E)$, in which
\begin{equation}
\begin{aligned}
\gamma_{A}(E)=\lim_{L\rightarrow\infty}\frac{1}{L}\ln\prod_{j=1}^{L}\frac{1}{\left|1-e^{-q}e^{i(2\pi\alpha j+\theta)}\right|}=\left\{\begin{matrix}
0,  & q\ge0, \\
-q, & q<0.
\end{matrix}\right.
\end{aligned}
\end{equation}
and $\gamma_B(E)$ can be calculated by Avila global theory~\cite{AAvila2015}. We insert an additional imaginary phase $\theta\rightarrow\theta+i\epsilon$. Although $T_{j}\in SL(2,\mathbb{C})$ is not an even function for $\epsilon$, we can handle it by complexification as well~\cite{YCZhang2022,YWang2021,YWang2020a,YWang2021a,AAvila2015}. Then, by letting $\epsilon\rightarrow-\infty$, one can get
\begin{equation}\label{Bj2}
B_{j,\epsilon}=e^{-q}e^{\epsilon}e^{i(2\pi\alpha j+\theta)}\begin{bmatrix}
-(E+\lambda e^{q}) & 1 \\ 
-1 & 0
\end{bmatrix}+\mathcal{O}(1).
\end{equation}
Then, we obtain $\gamma_{B,\epsilon}=\gamma_{B,0}+\epsilon+\mathcal{O}(1)$, where $\gamma_{B,0}=\ln\left|\frac{|E+\lambda e^{q}|+\sqrt{|E+\lambda e^{q}|-4}}{2}\right|-q$. Combined with $\gamma_{A}$, we obtain $\gamma_{\epsilon}=\gamma_{A}+\gamma_{B,0}+\epsilon+\mathcal{O}(1)$. Note that, $\gamma_{\epsilon}$ is a convex, piecewise linear function of $\epsilon$ with their slopes being integers. If the energy belongs to the spectrum, one can obtain the LE for $q>0$, i.e.,
\begin{equation}
\gamma=\ln\left|\frac{|E+\lambda e^{q}|+\sqrt{|E+\lambda e^{q}|-4}}{2}\right|-q.
\end{equation}
The critical energy of MEs satisfies $\gamma=0$. Then, one can obtain
\begin{equation}\label{mmee}
E_{c}=(1-\lambda) e^{q}+e^{-q}.
\end{equation}
The numerical fractal dimension $\Gamma$ as a function of $\lambda$ is plotted in Fig.~\ref{F3}(a), where the red dashed line is the exact ME $E_{c}$ from the expression Eq.~\eqref{mmee}. The numerical results are in good agreement with analytical expression. 

From the Fig.~\ref{F3}(a), as the quasiperiodic strength $\lambda$ increases, the first phase transition occurs at $E=2$, and the last phase transition occurs at $E=-2$ ($\lambda_c\approx3.5$). After that, the system enters the localized phase completely. For the case of $-2<E<2$, the critical points are in the interval $0<\lambda_c<3.5$. 

Furthermore, we can fix the base points $E_{b}=\pm 2$ and compute the winding number $\omega$ as a function of $\lambda$ to characterize the appearance of MEs and the complete localization transition, respectively [see the insets of Fig.~\ref{F3}(b)]. One can see that when the MEs appear, the $\omega$ for the base $E_{b} = -2$ transforms from 0 to 1 and the system has a topological point gap. When $\omega$ for the base point $E_{b}=\pm 2$ are both non-zero, the system enters the localized phase completely. Specifically, we show in the inset the fractal dimension $\Gamma$ in the complex plane under $\lambda = 1$ (coexistence of extended and localized states) and $\lambda=4$ (localized phase), which agree with our previous analysis well. Therefore, for the case of $N =\infty$, the localized region encroaches on the extended region from the maximum real energy, and there is always only one topological gap in the system. The corresponding maximum winding number $\omega=1$ (see Appendix ~\ref{B} for details about phase diagrams versus $N$). 

\section{The point gap in dual space}\label{dualspace}
From the previous section, one can find that the emergence of winding numbers and multi-loop nested point gaps is due to the superposition of different terms of the GSM potential. The analysis in dual space, one can further understand the physical meaning of point gaps.

\begin{figure}[tbp]
\centering
\includegraphics[width=8.5cm]{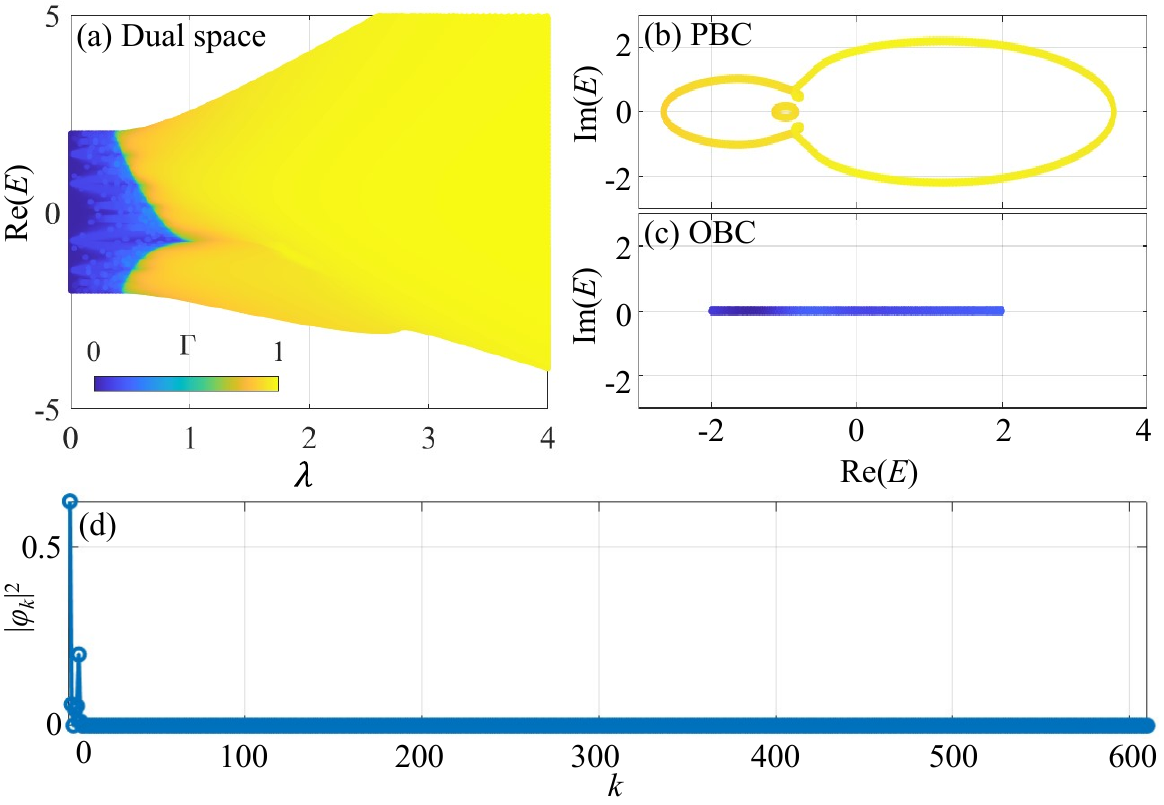}
\caption{(a) The dual space phase diagram. Eigenenergies in the complex plane with $\lambda=1.7$ under PBCs (b) and OBCs (c). (d) The dual space wave function of the eigenstate corresponding to an eigenenergy of $-0.9217$. The other parameters $L=610$, $\alpha=377/610$, $q=0.2$ and $N =2$.}
    \label{k1}
\end{figure}

By applying the dual transformation, one can get the dual space Hamiltonian, i.e.,
\begin{equation}\label{dual}  H_{k}=\sum_{n=1}^{N}\sum_{k=1}^{L-n}{\lambda}e^{-(n-1){q}}c_{k}^{\dagger}c_{k+n}+\sum_{k=1}^{L}2J\cos(2{\pi}{\alpha}k)c_{k}^{\dagger}c_{k},
\end{equation}
where a set of unidirectional long-range hopping terms emerge in the expression~\eqref{dual}. The parameter $q$ represents the decay strength of hopping terms, and the parameter $N$ determines both the number of hopping terms and the limit distance of the long-range hopping. We plot the dual space phase diagram in the $\lambda-E$ plane based on the fractal dimension property [see Fig.~\ref{k1}(a)]. The blue (yellow) region corresponds exactly to the localized (extended) phase, which means that the localization property in the dual space is opposite to that in lattice space. Furthermore, we exhibit the eigenenergy with $\lambda=1.7$ under PBCs [see Fig.~\ref{k1}(b)]. Consistent with the above conclusions, the eigenstates in the dual space in this case behave as extended states, which are exactly the opposite of the localized states in the lattice space.

On the other hand, directly from the mathematical structure of the dual space Hamiltonian~\eqref{dual}, one can find that there exist nonreciprocal hopping terms, which will lead to the emergence of skin effects in the dual space. This is another visual angle for topological point gaps.

Furthermore, we calculate the eigenspectrum under OBCs [see Fig.~\ref{k1}(c)]. The eigenspectrum completely collapses into the real axis, i.e., all the eigenvalues in the spectrum are pure real. The corresponding fractal dimensions reveal that all the states of the system are localized. As the skin mode, the corresponding wave function exhibits the characteristic of a left skin mode [see Fig.~\ref{k1}(d)]. In other words, the spectral winding number of the model can be used to describe the skin effect in the dual space.
\\
\section{Conclusion}\label{Sec.6}
In summary, we study localization properties and topological point gaps in a GSM quasiperiodic lattice. The results reveal that multiple MEs and point gaps with high winding number can emerge under the condition of summation terms' number $N>1$. The system's PT symmetry is damaged synchronously with the topological phase transition. Specifically, for the case of finite summation number $N$, the circle number of topological point gaps can at most be $N$. As the eigenstates continuously transform into localized states, the different point gaps gradually merge, and the maximum winding number gradually increases from $\omega=0$ to $\omega=N$. For the limit case of $N=\infty$, however, Avila's global theory can give the exact ME's expression. Analytical and numerical results consistently show that the corresponding MEs (topological point gaps) will merge together, resulting in a winding number $\omega=1$. Furthermore, the main results confirm that the properties of MEs and point gap structures are consistent with the behavior of topological boundaries and PT symmetry breaking. In principle, the corresponding phenomena can be realized by artificial quantum systems. As an example, we propose an experimental scheme based on the Rydberg atomic array~(see~Appendix~\ref{APPF}).

\section{Acknowledgements}
We thank R.-J. Chen, X.-D. Hu and W.-L. Li for their insightful suggestions. This work was supported by the National Key Research and Development Program of China (Grant No.2022YFA1405300), Open Fund of Key Laboratory of Atomic and Subatomic Structure and Quantum Control (Ministry of Education) and the Guangdong Basic and Applied Basic Research Foundation (Grants No.2021A1515012350).

\appendix

\section{Details for GSM lattice model}
To make a better understanding of the model, we provide deduction details. The Hamiltonian reads
\begin{equation}
\begin{split}H=&\sum_{j=1}^{L-1}(Jc_{j}^{\dagger}c_{j+1}+\mathrm{H.c.})\\&+\sum_{j=1}^{L}\sum_{n=1}^{N}{\lambda}e^{-(n-1)q}e^{in(2{\pi}{\alpha}j+{\theta})}c_{j}^{\dagger}c_{j},
\end{split}
\end{equation}
where
\begin{equation}
    V_{j}=\lambda e^{i(2\pi\alpha j+\theta)}
\end{equation}
for $N=1$;
\begin{equation}
V_{j}={\lambda}e^{i(2{\pi}{\alpha}j+{\theta})}+\lambda e^{-q}e^{2i(2{\pi}{\alpha} {j}+\theta)}
\end{equation}
for $N=2$;
\begin{equation}
V_{j}={\lambda}e^{i(2{\pi}{\alpha}j+{\theta})}+{\lambda}e^{-q}e^{2i(2{\pi}{\alpha}j+{\theta})}+{\lambda}e^{-2q}e^{3i(2{\pi}{\alpha}j+{\theta})}
\end{equation} 
for $N=3$.

The terms number of GSM potential increases with an increasing $N$. The potential forms a geometric sequence. Interestingly, as $N\rightarrow\infty$, the terms number can reduce to be one by using the geometric sequence formula. Then, one can obtain
\begin{equation}\label{V_jterm}
V_{j}=\frac{\lambda{e}^{i(2{\pi}{\alpha}j+{\theta})}-\lambda{e}^{-{N}q}{e}^{({N}+1)i(2{\pi}{\alpha}{j}+{\theta})}}{1-{e}^{-q}e^{i(2{\pi}{\alpha}j+{\theta})}}.
\end{equation}
Note that, when $N\rightarrow\infty$, the term $\lambda{e}^{-{N}q}{e}^{({N}+1)i(2{\pi}{\alpha}{j}+{\theta})}$ tends to $0$. then, one can obtain
\begin{equation}\label{V_j_}
V_{j}=\frac{{\lambda}e^{i(2{\pi}{\alpha}j+{\theta})}}{1-e^{-q}e^{i(2{\pi}{\alpha}j+{\theta})}}.
\end{equation}
Thus, the terms number is one and the corresponding maximum winding number reduces to $\omega=1$. By the way, Eq.~\eqref{V_j_} is a non-Hermitian version of the Ganeshan-Pixley-Das Sarma potential~\cite{SGaneshan2015}.
\\

\begin{figure*}[htpb]
\centering
\includegraphics[width=15cm]{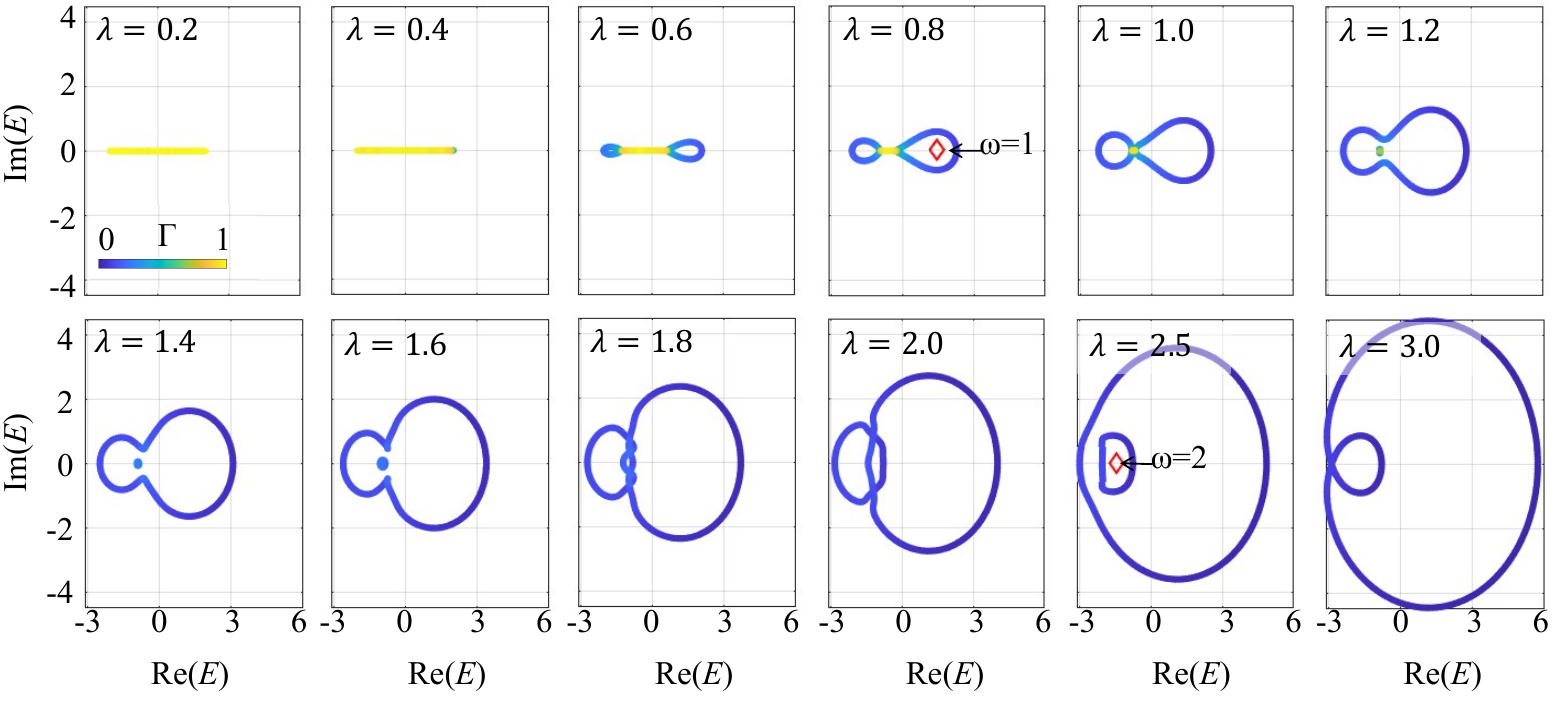}
\caption{The detailed evolution of point gap and the corresponding winding number with respect to $\lambda$ under the condition of $N=2$. The quasiperiodic strengths are marked. The other parameters $L=610$, $\alpha=377/610$ and $q=0.2$.}
    \label{F6}
\end{figure*}

\section{The evolution of point gap with respect to $\lambda$}\label{A}
\begin{figure*}[tbp]
\centering
\includegraphics[width=15cm]{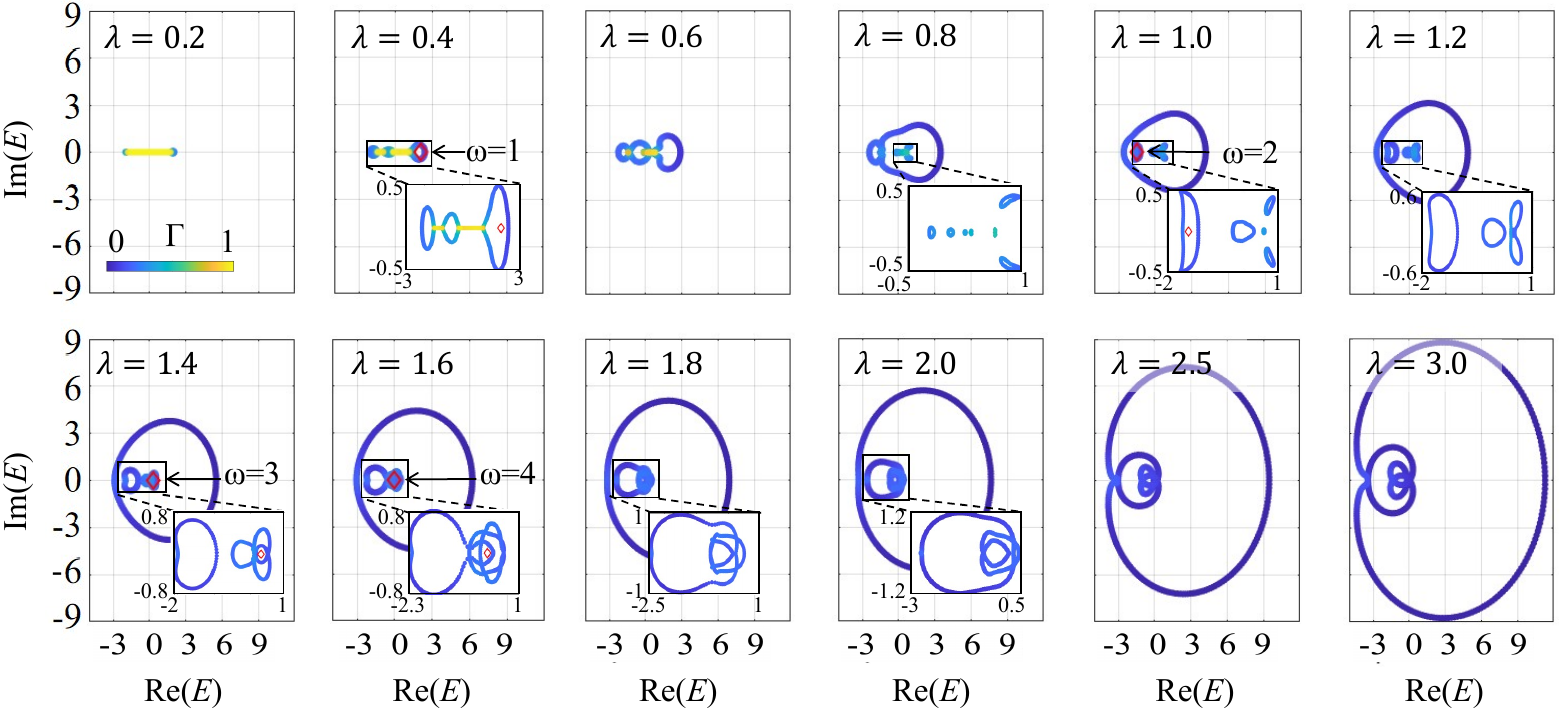}
\caption{The detailed evolution of point gap and the corresponding winding number with respect to $\lambda$ under the condition of $N=4$. The quasiperiodic strength are marked. The other parameters $L=610$, $\alpha=377/610$ and $q=0.05$.}
    \label{F7}
\end{figure*}

To exhibit the multi-loop nested point gaps and the process of finding the base point $E_b$, we provide two examples, i.e., the case of $N=2$ and $N=4$.

First, let's discuss the case of $N=2$. It is not difficult to find from Fig.~\ref{F6} that with the increase of quasiperiodic strength, some states with the largest eigenenergy in the system undergo localization phase transitions first. Then, as $\lambda$ continues to increase, some eigenstates with the lowest energy begin to enter the localized state. Then, the number of localized states will continue to increase with an increasing $\lambda$, and eventually all extended states will transform into localized states. Then, the two points will merge together to form a two-layer multi-loop nested structure. After that, by further increasing $\lambda$, the size of the ring become even larger, and the two-layer circle structure will always exist.

Now we turn to the case of $N=4$. With the increase of quasiperiodic strength $\lambda$, the pace of phase transition is similar to the former case, i.e., the states in the two ends of eigenenergy localized first, and the states in the middle will follow suit. Besides, unlike the $N=2$ case, there are more point gaps with high winding number. Specifically, there are point gaps with winding number $\omega=1$ (e.g., subfigure of $\lambda=0.4$), $\omega=2$ (e.g., subfigure of $\lambda=1.0$), point gaps of $\omega=3$ (e.g., subfigure of $\lambda=1.4$), $\omega=4$ (e.g., subfigure of $\lambda=1.6$)~[see Fig.~\ref{F7}].

\section{Discussions on phase diagram of different $N$}\label{B}
\begin{figure}[tbp]
\centering
\includegraphics[width=8.5cm]{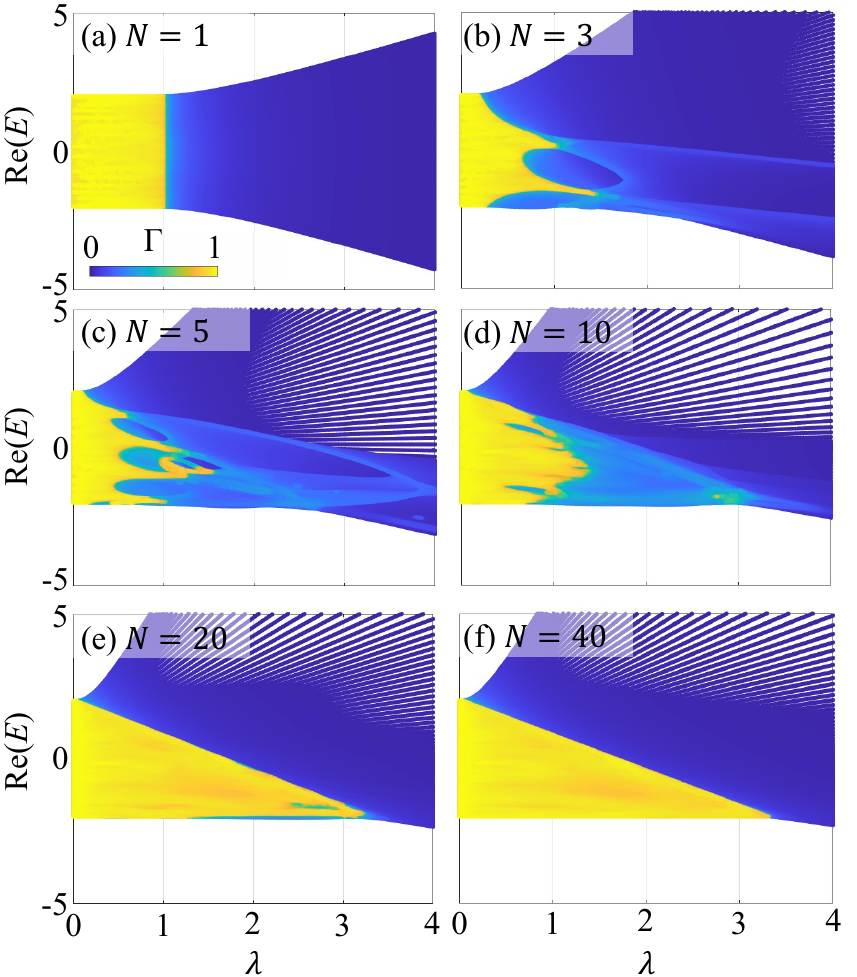}
\caption{$\Gamma$ phase diagram on $\lambda-E$ plane with different $N$. Parameters $\lambda=1.9$, $L=610$, $\alpha=377/610$ and $q=0.2$.}
    \label{F5}
\end{figure}

In fact, while the number of MEs increases with an increasing $N$, the interval between MEs decreases versus $N$. Then, for the case of $N\rightarrow\infty$, the interval between two nearest-neighboring MEs approaches $0$, and finally merges into one ME~[see Fig.~\ref{F5}]. 
The relationship between the ME and the parameter $N$ can be summarized as follows. Under the condition of $N=1$, the system exhibits a traditional topological point gap, in which there is no ME and only a standard critical point, namely $\lambda_c=J$. With the increase of parameter $N$, multiple MEs gradually appear in the system, and the number of MEs is $2(N-1)$, which is consistent with the previous prediction. Finally, as $N$ increases, the exponential decay term at large $N$ makes the MEs all merge together, while only one ME can be observed on the phase diagram. One can obtain the same conclusion from the point of view of topological point gap analysis, i.e., although the increase of point gaps' circle promotes a higher winding number with an increasing $N$, the maximum winding number decreases to $1$ due to the merging of point gaps' circles.

\section{Scaling effect}
\begin{figure}[tbp]
\centering
\includegraphics[width=8.5cm]{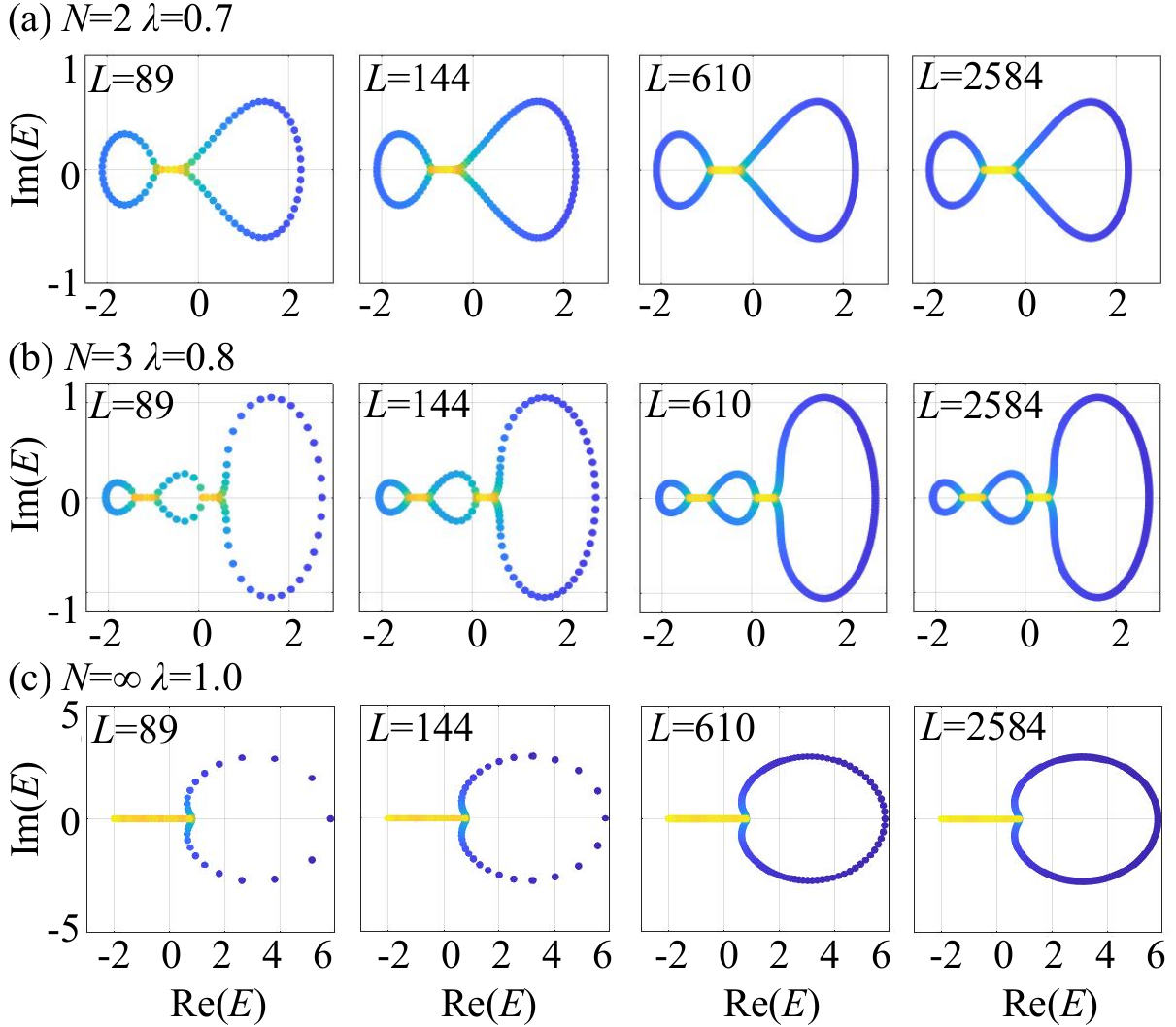}
\caption{The finite size scaling of point gap with $N=2$ (a), $N=3$ (b) and $N=\infty$ (c). The system size $L$ are marked. The corresponding $\alpha=\frac{55}{89},~\frac{89}{144},~\frac{144}{233},~\frac{377}{610}$ and $\frac{1597}{2584}$ respectively. Throughout, $q=0.2$.}
    \label{F8}
\end{figure}

According to the definition Eq.~\eqref{Gamma}, the fractal dimension is related to the system size. To ensure the correction, this section provides the result of finite size scaling in Fig.~\ref{F8}. As shown in the figure, the system size affects the points density of the loop structure in the complex energy plane, but does not qualitatively affect the localization and topological characteristics of the system. In concrete terms, the point distribution is sparse due to the small size~[see the case of $L=89$]. The energy spectrum is smoother and more complete with increasing calculation points. The calculation accuracy increases accordingly. Once the size exceeds $L=610$, neither the smoothness nor the completeness of the image undergoes further changes. Thus, $L=610$ is the optimal choice considering computing power constraints.

\begin{figure}[btp]
\centering
\includegraphics[width=8.5cm]{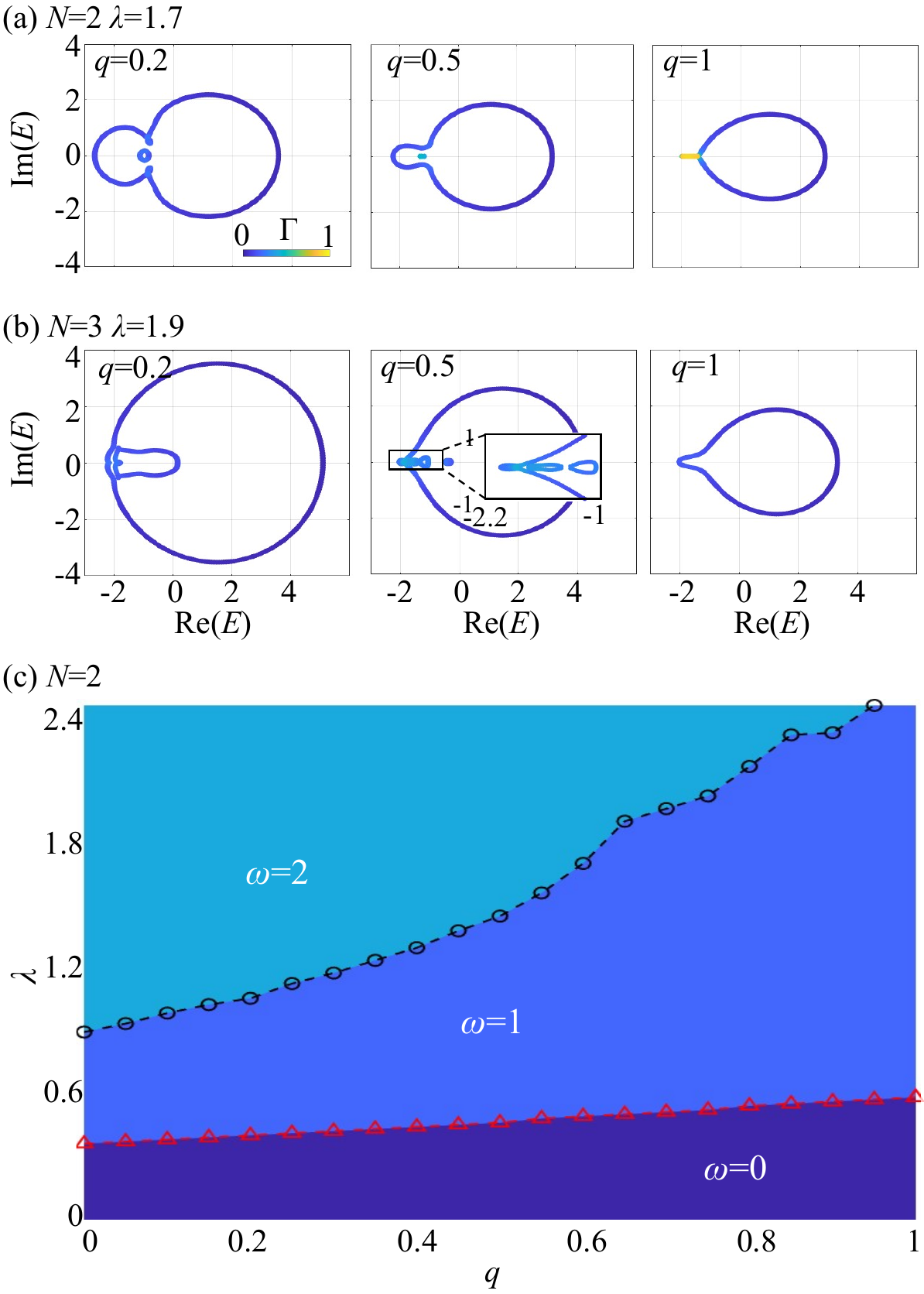} 
\caption{The point gap with $N=2$ (a) and $N=3$ (b) tuned by parameter $q$. (c) The phase diagram of winding number ${\omega}$ in the $\lambda-q$ plane when $N=2$. The other parameters $\alpha=377/610$ and $L=610$.}
 \label{F9}
\end{figure}

\begin{figure*}[htbp]
\centering
\includegraphics[width=12cm]{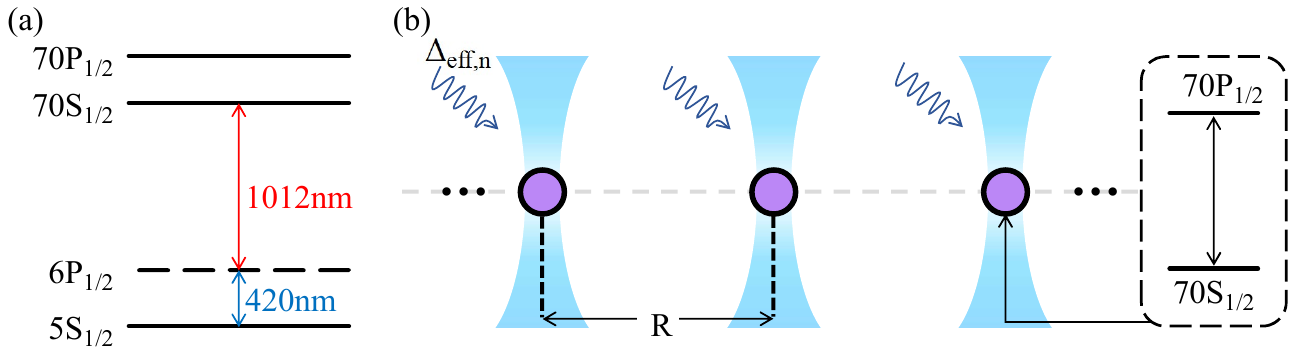} 
\caption{Experimental scheme on GSM model in a Rydberg atomic array. (a) Schematic diagram of atomic energy levels. (b) Rydberg atoms trapped in optical tweezers to form the atomic array.}
 \label{F10}
\end{figure*}

\section{The effect of $q$}
The parameter $q$ acts on the exponent of Eq.~\eqref{Vj} and plays an important role in modulating the quasiperiodic strength. In concrete terms, $q\to\infty$, $e^{-(n-1)q}\to 0$. On the other word, when $q$ increases, the quasiperiodic strength vanishes and the system becomes trivial regardless of the parameters adjusted. Only the metallic phase is presented, whereas the Anderson localization disappears. Therefore, the system becomes trivial for a large $q$. So, we consider $q\in(0,1)$ in our discussion. We provide the cases of $N=2,3$ as examples~[see Fig.~\ref{F9}]. When $q$ increases to $1$, extended states emerge from the system whose states were originally all localized~[see Fig.~\ref{F9}(a)]. Accordingly, the topological phase of the system gradually changes from $\omega=2$ to $\omega=1$. The results of $N=3$ are similar~[see Fig.~\ref{F9}(b)]. 

Furthermore, we calculate a phase diagram of the winding number $\omega$ in $\lambda-q$ plane. The system undergoes a topological phase transition with an increasing $\lambda$ [see Fig.~\ref{F9}(c)]. More importantly, the topological phase transition point is delayed as $q$ increases. The system fails to achieve $\omega=2$ until $q$ reaches a certain threshold. This is consistent with the previous results. For the purpose of effective discussion, we set $q=0.2$.

\section{Experimental realization and measurement}\label{APPF}
The model we propose in this paper is an extended version of the AA model. So far, the standard AA model has been realized in a variety of artificial systems, such as optical lattice ultracold atoms, ion traps, optical waveguide arrays, superconducting lines, Rydberg atomic array, and so on. The manipulation techniques of quasiperiodic disorder in these artificial quantum simulators are very mature. Based on the existing experimental platforms of AA model, it only needs to calculate the specific GSM potential at the corresponding lattice sites by strictly following the potential function expression given in this paper, i.e., Eq.~\eqref{Vj} in the main text.

Following this, as an example, we discuss the relevant experimental scheme on the $^{87}$Rb Rydberg array. Since the Rydberg atomic array corresponds to a spin model, we need to map the Rydberg atomic array to the model described by Eq.~\eqref{Hami}. Experimentally, the Hamiltonian required for the Rydberg experiment is given by
\begin{equation}\label{spinHami}
H_{s}=\sum_{j_{x}}( J\sigma_{j_{x}}^{+}\sigma_{j_{x}+1}^{-}+H.c.)+\frac{1}{2}\sum_{j_{x}}V_{j_{x}}(\mathbb{I}+\sigma_{j_{x}}^{z}),
\end{equation}
where $\sigma^{\pm}=\frac{1}{2}(\sigma_{x}\pm i\sigma_{y})$ with Pauli matrices $\sigma_x$ and $\sigma_y$. The Hamiltonian~\eqref{spinHami} can be transformed to the GSM lattice by defining operator $c^{\dagger}_j=\sigma^{+}_{j_{x}}=\left|\uparrow\right>_{j_{x}}\left<\downarrow \right|_{j_{x}}$ at each site $j_{x}$, where $\left|\uparrow\right>= 70P_{1/2}$ and $\left|\downarrow\right>= 70S_{1/2}$ for $^{87}$Rb atoms~[see Fig.~\ref{F10}]. The dipole-dipole interaction between Rydberg atoms is given by $J= \frac{d^2}{R^3}$, where $d$ represents the transition dipole moment between the two Rydberg levels, $R$ is the distance between two nearest neighboring sites, $70$ is the principal quantum number, and the energy unit $J=2\pi\times7.53$MHz. Note that, since the non-nearest neighboring effect decays with distance $R^{\nu}$ ($\nu\geq3$), one can safely ignore the corresponding terms.

Specifically, the experiment on Rydberg atomic array includes four steps. The first step is to load the atoms into the optical tweezers. The second step is to excite the atoms to the Rydberg state of the principal quantum number $70$. The third step is to introduce the GSM potential energy into the atomic array. The fourth step is to measure the eigenvalues and eigenstates of the system and then calculate the corresponding fractal dimension. A detailed introduction will be provided below.

At first, optical tweezers, which are generated by a spatial light modulator (SLM), will be used to form an atomic chain of $^{87}$Rb atoms~[see Fig.~\ref{F10}]. Specifically, one can capture the atoms by optical tweezers, and then rearrange them by an acousto-optical deflector (AOD). Consequently, one can obtain a one-dimensional array of $^{87}$Rb atoms.

Secondly, the $^{87}$Rb atoms will be excited from the $5S_{1/2}$ state to the $70S_{1/2}$ state. One can achieve it by coupling the $5S_{1/2}$ and $70S_{1/2}$ states through 420nm and 1012nm lasers, with the $6P_{1/2}$ state acting as the intermediate state during this two-photon process.

Thirdly, by irradiating a 1012nm Rydberg laser to the corresponding atoms~(with a detuning of about 100MHz to the 1012nm Rydberg laser) and considering the decoherence~\cite{TMWintermantel2020}, different sites can achieve an on-site potential, which satisfies the quasiperiodic properties. This potential is given by $V_{AC}=V_{on-site}=\frac{|\Omega_0|^2}{\Delta}[\frac{1}{2}\sum_{n=1}^{N}e^{-(n-1)q}e^{in(2{\pi}{\alpha}j_{x})}]^{2}$, where $\Omega_0$ is the corresponding laser Rabi frequency. Specifically, this process can be carried out by introducing a second spatial light modulator (SLM). By generating a locally controllable light shift, one can achieve site-dependent detuning $\Delta_{\text{eff},n}$, as marked in Fig.~\ref{F10}(b). The corresponding method has recently been reported by the Harvard group~\cite{TManovitz2024}.

At last, one can use a dynamical method to measure the corresponding phenomena on the Rydberg array~\cite{PRoushan2017, SZLi2025}. In concrete terms, one can excite a Rydberg atom to a superposition state of $70S_{1/2}$ and $70P_{1/2}$ as the initial state, and then let the system evolve. The eigenvalues and the corresponding density distributions can be obtained by Fourier transform.

\end{document}